# A Continuous Beam Steering Slotted Waveguide Antenna Using Rotating Dielectric Slabs

Amirhossein Ghasemi and Jean-Jacques Laurin

*Abstract*—The design, simulation and measurement of a beam steerable slotted waveguide antenna operating in X band are presented. The proposed beam steerable antenna consists of a standard rectangular waveguide (RWG) section with longitudinal slots in the broad wall. The beam steering in this configuration is achieved by rotating two dielectric slabs inside the waveguide and consequently changing the phase of the slots excitations. In order to confirm the usefulness of this concept, a non-resonant 20-slot waveguide array antenna with an element spacing of $d = 0.58\lambda_0$ has been designed, built and measured. A 14° beam scanning from near broadside ($\theta = 4°$) toward end-fire ($\theta = 18°$) direction is observed. The gain varies from 18.33 dB to 19.11 dB which corresponds to the radiation efficiencies between 95% and 79%. The side-lobe level is -14 dB at the design frequency of 9.35 GHz. The simulated co-polarized realized gain closely matches the fabricated prototype patterns.

## I. Introduction

MANY types of beam scanning antennas have been designed and developed in the past. In particular, radar sensors for automotive applications use antennas with agile beams, either by mechanically moving the complete antenna or electronically scanning beam with the fixed antenna [1]. Electromechanical beam scanning offers a low cost solution but it is less agile compared to electronically steerable antennas. Moreover, it can potentially be used at higher peak power levels because no electronic parts are used [2]. Special attention was paid to concept of a fixed antenna in which small parts are moved. Many such concepts have been presented in [3-6], as candidates for millimeter wave scanning antennas and in [7-9] for satellite communications. The *Eagle Scanner* of World War II was possibly the first electromechanically beam scanning antenna and included displacement of the narrow wall of a rectangular waveguide [10] in order to introduce phase shifting in a series-fed linear array. However, the realization of precise and instantaneous linear motion of the side wall over the total length of the antenna is difficult to achieve.

Recent works on reconfigurable antennas have provided an alternative solution to perform the beam steering with less complexity [11]. Such reconfigurable antennas use active components such as PIN diodes [12-13], varactor diodes [14] and microelectromechanical systems (MEMS) [15] to control the beam steering/pattern of the antenna. A possible drawback of using RF PIN as a switch is that it requires additional passive elements for the DC biasing circuitry, which may affect the antenna dimensions and efficiency and also in the return loss

A. Ghasemi and J. J. Laurin are with Poly-Grames Research Center, École Polytechnique de Montréal, Montreal, QC H3C 3A7, Canada, (e-mail: Amirhossein.Ghasemi@polymtl.ca and Jean-Jacques.Laurin@polymtl.ca).

performance [11]. Moreover, as the power increases, implementing the active components becomes more difficult and challenging because of their nonlinear response.

Frequency scanning antennas based on series-fed arrays or leaky-wave structures may possibly offer a simple and low-cost solution compared with phased arrays. However, frequency variation over a wide bandwidth to scan the main beam is required [16] which is not suitable for the narrow bandwidths generally allocated to the certain applications such as X-band weather radars.

In order to simplify the requirements of the beam scanning mechanism, a new concept is proposed which consists of a waveguide linear slot array in which beam scanning is achieved by rotating of two dielectric slabs inside the rectangular waveguide. The configuration is inspired by an earlier design of a waveguide phase shifter presented in [17]. In this design, the wavelength of the travelling wave is changed depending on the angle of the slabs relative to the dominant mode field.

The lengths of the slots and their offsets from the waveguide center plane of the slots need to be determined such that a specified pattern and a specified input impedance level are achieved. To do this successfully, we must account for the mutual coupling between slots. One way to consider the mutual coupling is to simulate an array of several coupled slots and extract their characteristics. For radiating slots, design curve of resonant length and offset from the waveguide center plane against slot admittance are required. Once design data is obtained for the radiating slots, the next steps is to design a slot array antenna.

The proposed design allows continuous beam scanning without semiconductor devices and therefore can handle high power compared to the concepts presented in [12-13]. Although this concept has been first presented in [4] and worked out in [1], this work exhibits more scan range with low side-lobes due to symmetric field distribution inside the waveguide. Furthermore, a matched transition from a coaxial port to the waveguide filled by the rotating slabs is presented.

The paper is organized as follows. Section II introduces the design methodology of the slotted waveguide antenna. In Section III, engineering details on the design method are given. Finally, simulation and experimental results are provided and discussed in Section IV.

## II. Design of the Slotted Waveguide Antenna

### A. Rotating Dielectric Slabs inside the Waveguide

The experimental evidence of the scanning capabilities realized by rotating a single ridge in a metal waveguide was given by Solbach *et al.* [1]. However, using a metallic ridge inside a metal allows the TEM mode to propagate and it becomes difficult to separate the different modes at the operation frequency. In [1], another drawback was the presence of high side lobes due to the asymmetric field distribution inside the waveguide. The asymmetry of the structure can be observed in Fig. 1a, showing the non-symmetric excitation of the slots on the RWG when the position of single slab is oblique (45º). To avoid the asymmetric coupling, we propose using two symmetrically positioned rotating slabs in the waveguide, as shown in the Fig. 1b. Also, the slabs are made of dielectrics to eliminate the TEM mode. The dimensions of the waveguide and the dielectric slabs have to be chosen to ensure single mode propagation at the desired frequency of operation, and to maximize the variation of the mode propagation constant with the rotation of the slabs. In order to simplify the fabrication, was used a standard waveguide size, namely WR-62. We have chosen Rogers RT/duroid 6006 for the dielectric rotating slabs, with the dimensions of Table I. Fig. 2 shows the propagation constant versus frequency while the dielectric slabs are in three different rotation states. These curves were obtained with the 3D full-eave solver of Ansys-HFSS, with no slots on the waveguide. In comparison with an air-filled waveguide, one notes that the cut-off frequency decreases when a fraction of the guide is filled by dielectric. The cut-off frequency of the air-filled WR62 waveguide is 9.49 GHz. In Fig. 2 we see that it varies between 7.5 and 8.3 GHz for the fundamental mode, depending on the rotation state of the slabs. It can be noted that at the chosen operation frequency, 9.35 GHz applicable to weather radars, the 2nd mode is well attenuated and the waveguide supports only one mode for all the rotation states of the slabs.

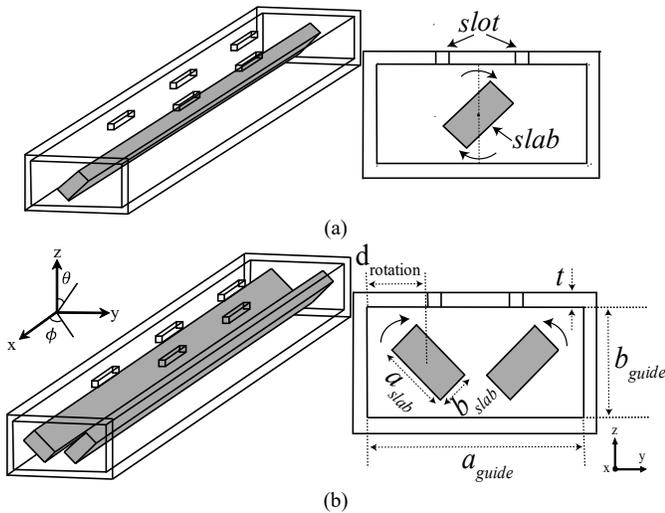

Fig. 1. Simulation model of the rotating dielectric (a) single slab, (b) two symetrically positioned slabs.

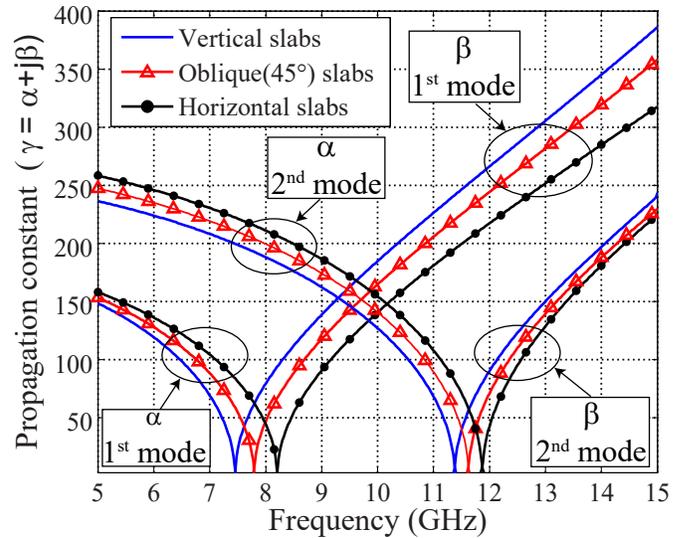

Fig. 2. Propagation constant for the different rotation states of the dielectric slabs. Waveguide simulated without slots.

TABLE I
Dielectric Slab and Waveguide Parameters

| $a_{guide}$ | $b_{guide}$ | $a_{slab}$ | $b_{slab}$ | t | $\varepsilon_{r\_slab}$ | $\tan \delta_{slab}$ | $d_{rotation}$ |
|---|---|---|---|---|---|---|---|
| 15.8mm | 7.9mm | 5mm | 2.5mm | 1mm | 6.15 | 0.0019 | 4.45mm |

B. *Non-Resonant Array with the Slots Alternatively Displaced*

The two basic types of slotted waveguide antennas are resonant array (standing wave) and non-resonant array (travelling wave). Since we wish to scan the beam off-broadside, the case of greater interest is the non-resonant array. For that case, the slot-to-slot spacing differs from $\lambda_g/2$ so the reflections from the different slots do not add up in phase at the input of the waveguide and the reflection coefficient will be small. Thus, the aperture distribution experiences a phase progression that is uniform, or nearly so, which is why these arrays are also referred to as travelling wave fed arrays [18]. Non-resonant arrays include a matched-load termination, necessary to avoid undesirable reflection causing a back-lobe responsible for the degradation of the antenna pattern. The main advantage of these arrays is a larger bandwidth in terms of side lobe level (SLL) and input matching, which makes them suitable for performing as beam scanning antennas [18]. In the non-resonant slotted waveguide antenna, the slot spacing can be chosen so that we can produce a main lobe at almost any arbitrary angle $\theta$ relative to the axis of the array. If offsets are alternated on opposite sides of the symmetry plane, then the array factor is given by [19]:

$$AF = \sum_{n=1}^{N} a_n e^{[jn(\beta_0 d \cos\theta - \beta_g d + \pi)]} \quad (1)$$

where $a_n$ is the slot excitation amplitude level, $d$ is slot spacing, $\beta_g$ is the guided phase constant and $\beta_0$ is the propagation constant in free-space. In order to have a good aperture efficiency, it is desired to scan the beam near broadside. Therefore, a non-resonant array with the slot spacing $d \neq \lambda_g/2$ is necessary. The element spacing of $d/\lambda_{g\_oblique(45°)\,slab} = 0.4$ was chosen because array factor calculations based on the $\beta_g$ values given in Fig. 2 for the design frequency of 9.35 GHz led to maximum beam deviation. By rotating the dielectric slabs from vertical to horizontal position, the wavenumber variation observed from Fig. 2 is

$$\beta_{g\_vertical\,slab} - \beta_{g\_horizontal\,slab} = 45.6 \text{ rad/m} \quad (2).$$

This gives 14° of beam scanning based on (1). A 20-slot waveguide was chosen in this study based on limitations of the fabrication means available to the authors. A triangular amplitude distribution superimposed on a constant lower level was used in order to demonstrate the capability to control the SLL.

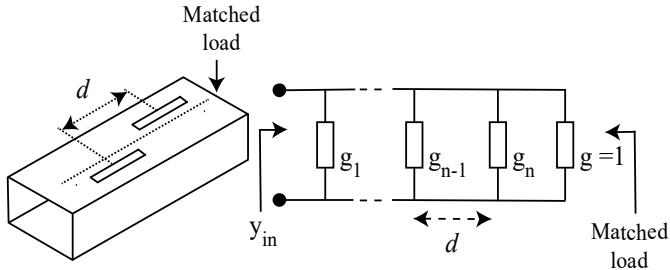

Fig. 3. Physical form and equivalent circuit model of the non-resonant array.

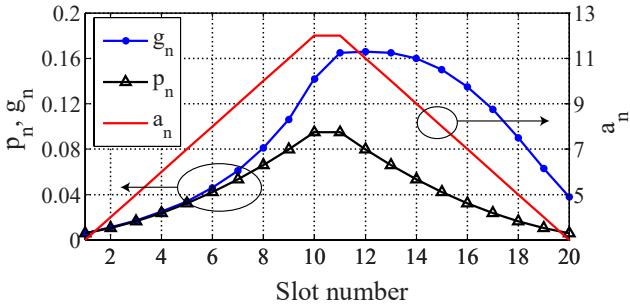

Fig. 4. Distribution of amplitude ($a_n$), radiated power ($P_n$) and normalized conductance ($g_n$) of the 20-element non-resonant array.

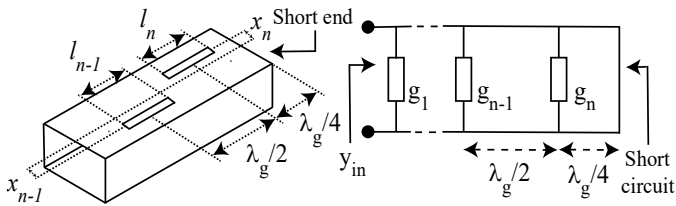

Fig. 5. Physical form and equivalent circuit model of the resonant array.

If the relative excitation level of the n$^{th}$ slot is $a_n$, the power $P_n$ radiated by this slot will be proportional to $a_n^2$. Thus, when we specify the required amplitude distribution $a_n$ to yield the desired beamwidth and side-lobe level we will know the $P_n$ within a constant of proportionality ($P_n = ka_n^2$) [19].

Let $r$ be the fraction of the incident power to be dissipated in the match load. The equivalent circuit for the array is shown in Fig. 3. Assuming a 1-watt power input and a lossless waveguide, we must have $r + \sum_{n=1}^{N} P_n = 1$. According to the non-resonant array design method presented in [19] we have $g_n = \frac{P_n}{r+\sum_{i=n}^{N} P_i} = \frac{P_n}{1-\sum_{i=1}^{n-1} P_i}$. Fig. 4 shows the values of distribution amplitude ($a_n$), radiated power ($P_n$) and normalized conductance ($g_n$) for a triangular taper excitation of a 20-element array, with $r = 0.15$.

### C. Evaluation of Offset and Length of the Slot

The design of the series-fed array is simpler if we can ignore the effect of coupling with adjacent elements. In order to observe if coupling has a significant effect, we have simulated resonant arrays of various length. Although we are considering scanning non-resonant arrays, the resonant array case is useful to estimate the admittance of the slots. As Fig. 5 illustrates, the slots of resonant arrays are spaced by $\lambda_g/2$, so that all of them share the same phase excitation necessary to produce a broadside main lobe. Typically, a short circuit is placed at the end of the waveguide, at a distance of $\lambda_g/4$ after the last slot. Assuming no coupling between the elements, the input admittance $y_{in}$ is the sum of all the admittances ($y_{in} = \sum y_i$) due to the spacing of $\lambda_g/2$ between the elements. Fig. 6 shows simulated values of $y_{in}$ for resonant slot arrays of 1 to 10 identical elements. Dielectric slabs with $\varepsilon_r = 6.15$ and $\varepsilon_r = 10.2$ were considered to illustrate the effect of the slab on coupling. The slabs were rotated at 45°. In both cases, a linear reference line was added using the admittance real part result for $N = 10$. Because of mutual coupling between the slots, especially for the $\varepsilon_r = 10.2$, the conductance curves do not follow the expected linear relationship. The best fit with the linear reference line is obtained for arrays of 3 to 6 identical slots with the dielectric of $\varepsilon_r = 6.15$. This dielectric constant causes less mutual coupling and is therefore a better candidate for this design.

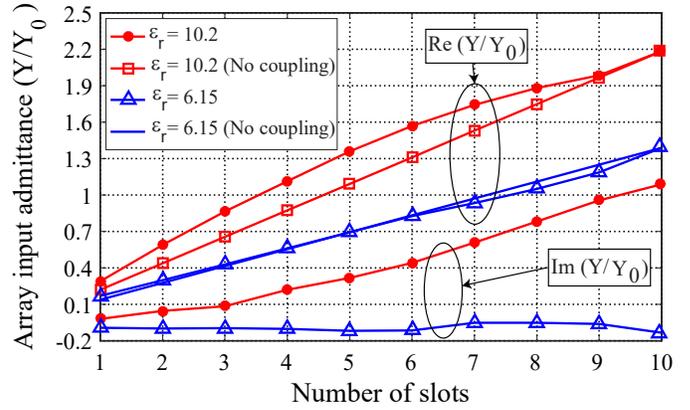

Fig. 6. The input admittance of the resonant array.

For each slot, we need to find the resonant length ($l_r$) for each given slot offset ($x_i$). This was accomplished by running a parameter sweep with HFSS for each $x_i$ in a 5-slot resonant array. The normalized susceptance ($b$) of the first slot must be zero at resonance. The parameter sweep returns a set of admittance points that was processed in order to determine $l_r$ (i.e. corresponding to zero susceptance) for a given slot offset.

For that, we used an interpolation routine implemented in Matlab to generate a contour corresponding to the resonant condition in the length-offset plane. This contour is shown in red in Fig. 7. Then, for each point on this contour we can determine the corresponding conductance ($g$) at resonance, see the dotted curve in Fig. 7. At the end of this process, two interpolation polynomials ((2) and (3)) were derived to give the slot resonant offset ($x_r$) and $l_r$ versus required resonant slot conductance as illustrated in Fig. 8.

$$x_r = p_1 g^2 + p_2 g + p_3 \tag{2}$$
$$l_r = q_1 g^3 + q_2 g^2 + q_3 g + q_4 \tag{3}$$

where:

$$p_1 = -0.1915, p_2 = 1.218, p_3 = 0.7608$$
$$q_1 = 0.2164, q_2 = -0.9546, q_3 = 1.571, q_4 = 14.76$$

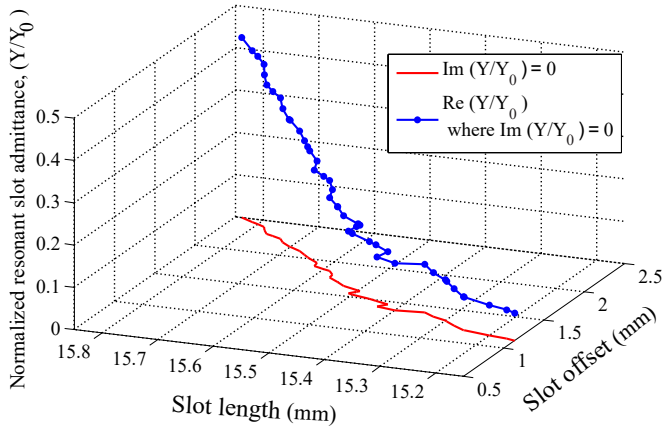

Fig. 7. Slot shunt admittance versus slot length and slot offset, (a) Conductance (Re( $Y/Y_0$ )), (b) Susceptance (Im( $Y/Y_0$ )).

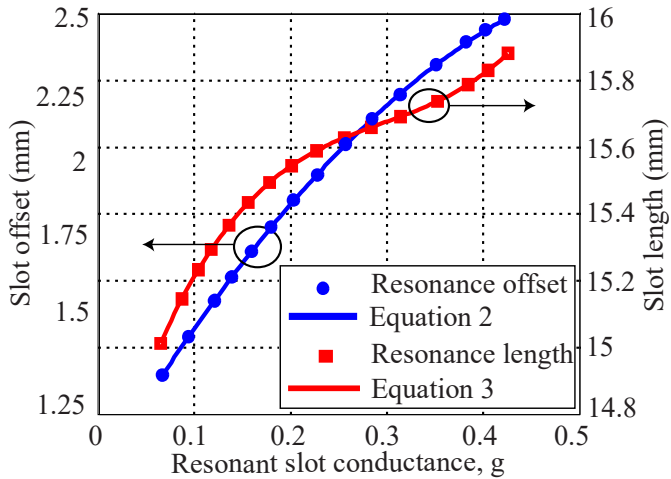

Fig. 8. Slot offset and slot length versus conductance (Re( $Y/Y_0$ )) obtained with curve fitting method for slabs rotation angle of $45°$.

TABLE II
Slot current distribution $a_n$, conductance $g_n$, offset $x_n$ and length $l_n$ for the 20-slot array design using triangular amplitude distribution

| n | $a_n$ | $g_n$ | $x_n$ (mm) | $l_n$ (mm) |
|---|---|---|---|---|
| 1 | 3 | 0.006 | 0.80 | 14.81 |
| 2 | 4 | 0.011 | 0.83 | 14.84 |
| 3 | 5 | 0.017 | 0.86 | 14.89 |
| 4 | 6 | 0.025 | 0.91 | 14.94 |
| 5 | 7 | 0.034 | 0.96 | 15.00 |
| 6 | 8 | 0.046 | 1.03 | 15.07 |
| 7 | 9 | 0.061 | 1.11 | 15.16 |
| 8 | 10 | 0.081 | 1.22 | 15.25 |
| 9 | 11 | 0.106 | 1.35 | 15.36 |
| 10 | 12 | 0.142 | 1.53 | 15.47 |
| 11 | 12 | 0.165 | 1.64 | 15.53 |
| 12 | 11 | 0.166 | 1.64 | 15.53 |
| 13 | 10 | 0.165 | 1.64 | 15.53 |
| 14 | 9 | 0.160 | 1.61 | 15.52 |
| 15 | 8 | 0.150 | 1.57 | 15.49 |
| 16 | 7 | 0.135 | 1.50 | 15.45 |
| 17 | 6 | 0.115 | 1.40 | 15.39 |
| 18 | 5 | 0.090 | 1.27 | 15.29 |
| 19 | 4 | 0.063 | 1.13 | 15.17 |
| 20 | 3 | 0.038 | 0.99 | 15.03 |

Using the values of conductance from the triangular distribution given in Fig. 3 and equations (2) and (3), we can obtain the resonant parameters of the antenna. Table II summarizes the final antenna parameters.

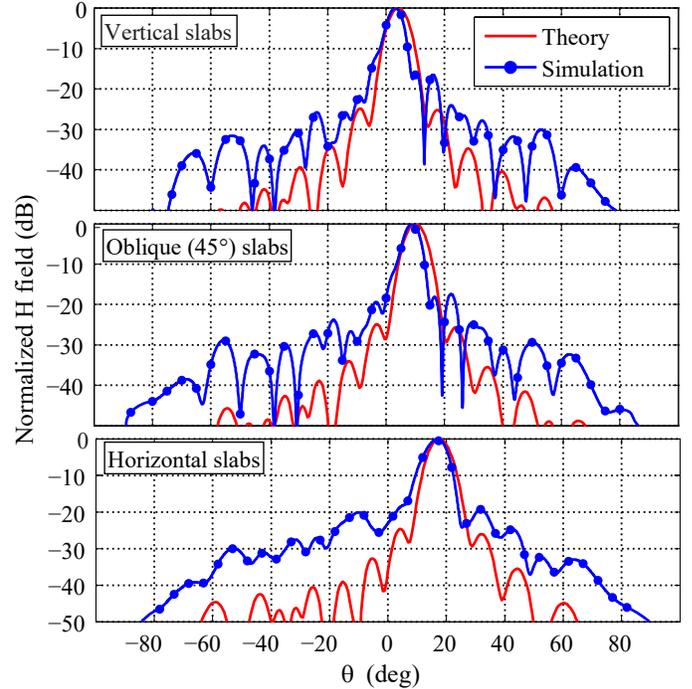

Fig. 9. Comparison of the normalized antenna factor; simulated H-plane (($\phi = 0$), see the axes presented in Fig.1) with HFSS with the theoretical array factor.

The antenna can then be simulated with HFSS.

### D. Array simulation

In order to validate the design approach, the values of $g_n$ in Table II were used in a circuit simulation (ADS® from Keysight) and the currents in the shunt conductances representing the slots were used to calculate the array factor using (1). The electrical length of the line sections between the shunt conductances with $\varepsilon_r = 6.15$ were calculated by HFSS. The resulting array factors are plotted in Fig. 9 (solid curves), along with the normalized H-plane patterns of the array with the various slabs rotation angles simulated with HFSS. The beam steering of 14° predicted by variation of $\beta_g$ is nearly achieved in both methods as can be seen in Fig. 9. Our method to determine the $g_n$ values is approximate because it assumed identical slots. Although mutual coupling is weak, as seen in Fig. 6, this effect was not taken into account in the design, which may explain the disagreement between the side lobe levels in the theoretical and simulated patterns. In addition, the scattering has occurred at the outer ends of the waveguide while in the theoretical model we do not have this effect.

## III. DESIGN GUIDELINES

### A. Matched load design

Non-resonant slotted waveguide arrays require a matched load termination. In this section, the design of a matched load for the dielectric loaded waveguide used in the antenna is proposed. Standard rectangular waveguide matched load cannot be used since our waveguide is loaded with dielectric slabs. However it is quite straighthforward to design a matched load by inserting absorber material. At this end, we have chosen the DD-10214 Silicon from ARC Technologies Inc, with $\varepsilon = (17 - 0.2j)\varepsilon_0$ and $\mu = (1.6 - 1.8j)\mu_0$ at the operation frequency of 9.35 GHz. Absorber strips with the thickness of $0.762_{mm}$ were placed on the broadwall of the waveguide. Based on this information, a load for WR62 with rotating dielectric slabs has been designed. As illutrated in Fig. 10, a stand made of teflon was used to support the slabs and since it is located after the absorber, it is not causing reflections. Return loss greater than 25 dB was obtained in simulations over a wide frequency range for the all rotation angles of the dielectric slabs (see Fig. 10).

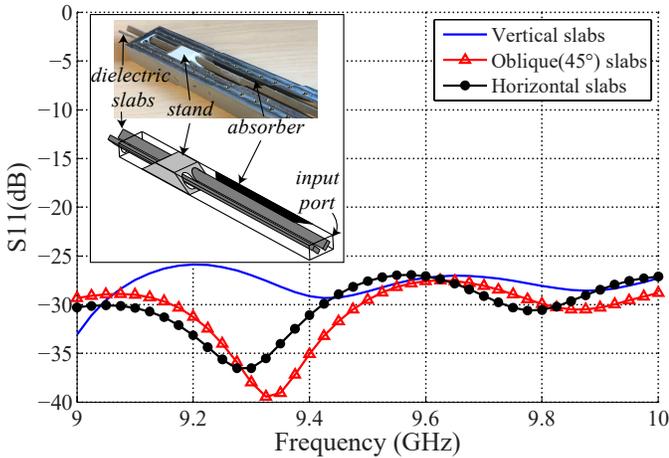

Fig. 10. WR62 matched load: simulated model, prototype and return loss simulation result in different slabs positions.

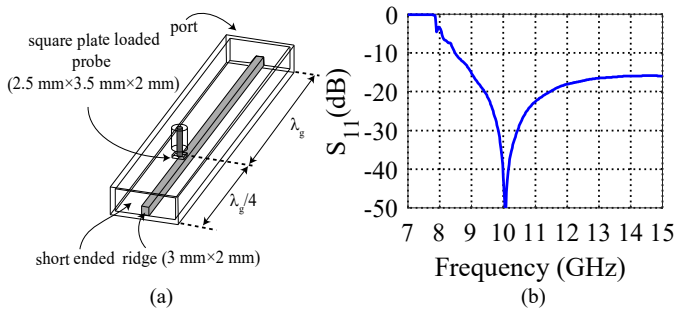

Fig. 11. (a) Coaxial to waveguide transition, (b) simulated return loss.

*B. Waveguide transitions*

The cut-off frequency of the air-filled WR62 waveguide is 9.49 GHz. Since our operation frequency for the RWG loaded with dielectric slabs is 9.35 GHz, it is required to reduce the cut-off frequency of the waveguide feeding the antenna. This was realized with a ridge waveguide, as shown in Fig. 11, which reduced the cut-off frequency to 7.8 GHz. A coaxial to ridge waveguide transition was designed. The size of the coaxial probe and its location with respect to the shorting wall were varied to optimise matching. Simulated return loss of more than 20 dB was achieved in the frequency band of interest, as shown in Fig. 11. Then, a second transition, this time from the ridge waveguide to a waveguide loaded with dielectric slabs was designed. Wedge shaped tips on both the ridge and the dielectric slabs led to good simulated return loss for all rotation angles of the slabs, as shown in Fig. 12 for three cases.

*C. Directional flare*

In order to improve the gain of the antenna, a low-profile flare section was added on the broad wall radiating surface of the waveguide (Fig. 13). The dimensions of the directional flares were varied in HFSS simulations in order to achieve maximum antenna gain. The optimal values of $H$ and $\alpha$ were shown in Fig. 14. The realized gain increases by about 5 dB. It should be noted that, if a much larger value of H is used, a horn is formed and the gain will increase more but the 3-dB gain beamwidth in the E-plane will decrease which means the antenna would not suitable for providing a fan beam.

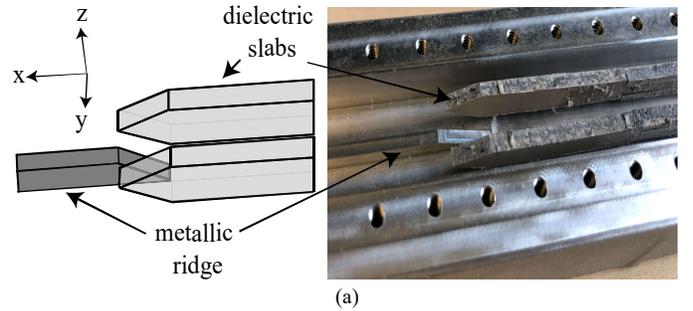

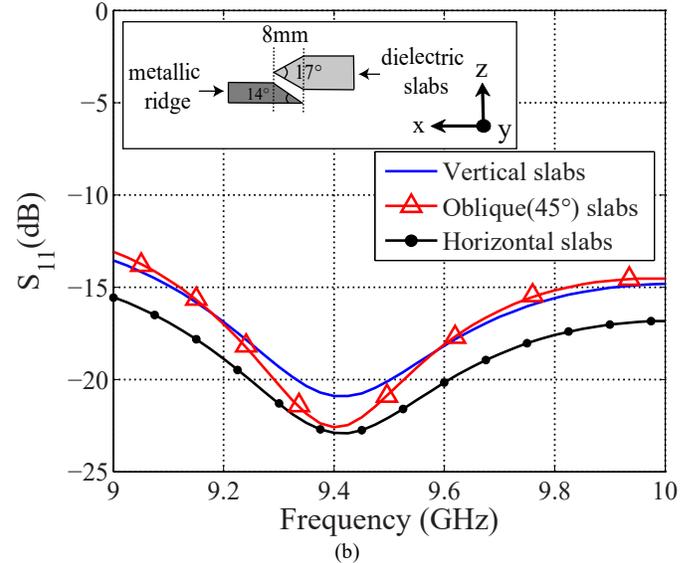

Fig. 12. Triangular tapering cross section of the dielectric slabs and the ridge, (a): Simulated model and fabricated prototype, (b): Return loss Coaxial to waveguide transition.

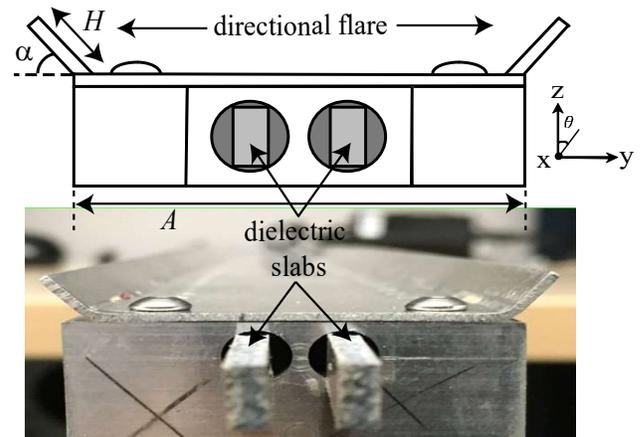

Fig. 13. Configuration of the waveguide with directional flare. The parameters are $A = 32$mm, $H = 5.5$mm, and $\alpha = 40°$.

IV. PROTOTYPE OF THE ANTENNA AND MEASUREMENT

An antenna prototype has been built using the WR62 standard waveguide according to the detailed design given in sections II and III. The 20 slots have been milled onto the upper broad wall in accordance with the dimensions of the resonant parameters given in Table II. The coaxial to waveguide transition and the matched-load described in Section IV were implemented. A prototype of the antenna can be seen in Figure 15. The dielectric slabs can be seen at the end of the waveguide. In our test they were rotated manually but adding a controlled motor is an easy task. It should be noted that if the antenna is designed for continuous beam steering with continuous rotating of the dielectric slabs, its performances have been measured for some rotating states from vertical to horizontal with a step of 45°.

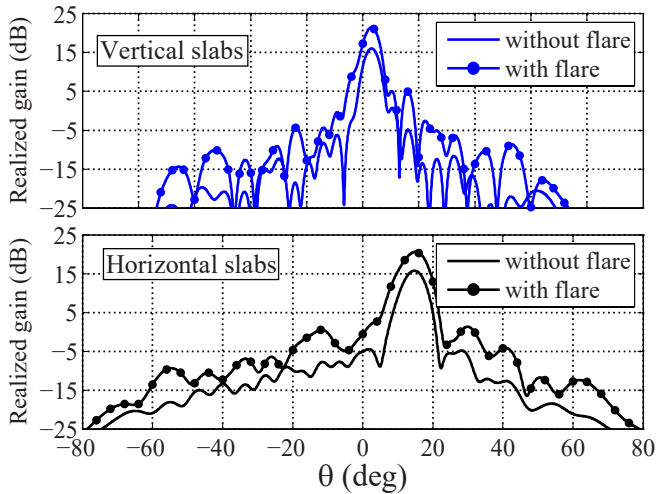

Fig. 14. Effect of the flare on the simulated H-plane radiation pattern (($\phi = 0$), see the axes presented in Fig.1) of the antenna.

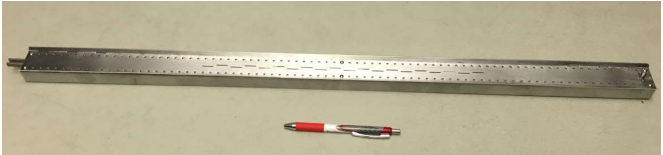

Fig. 15. Prototype of the 20-slot array antenna.

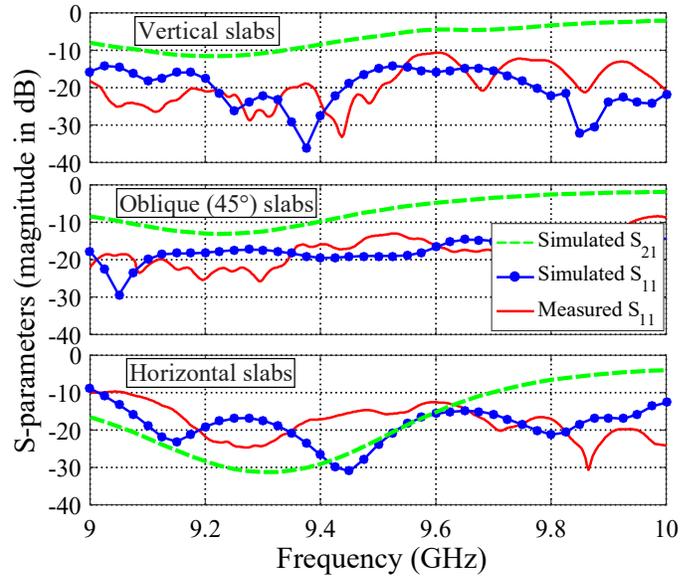

Fig. 16. Return loss of the antenna, simulated and measured in different position of the dielectric slabs.

### A. Impedance matching

Measurement of the return loss has been made on the antenna coaxial port for three positions of rotating slabs. The results are shown in Fig. 16, together with the simulations. The antenna was designed for a frequency of 9.35 GHz and a slabs rotation of 45°. We can see that return loss is greater than 10 dB for the three rotation states. A good agreement between measured and simulation $S_{11}$ parameters is obtained for three slab rotation states over the frequency band of 9 GHz to 10 GHz (BW = 10.5%).

The power lost in the terminating load was not measured due to the non-standard waveguide at the end of the antenna, but the $S_{21}$ parameters were simulated with HFSS and are also shown in Fig. 16. The array was designed by assuming dissipation of 15% (-8.2 dB) of the input power in the terminating load. We can see in Fig. 16 that the simulated $S_{21}$ at the design frequency of 9.35 GHz for the three rotation states is less than -10 dB which is lower than the assumption.

### B. Radiation patterns

The radiation properties of the antenna prototype were simulated with HFSS and have been measured in an anechoic chamber. H-plane patterns of the realized gain at the design frequency of 9.35 GHz are presented in Fig. 17. According to Fig. 17, a very good agreement is obtained between simulations and measurements for the three scanning states. As expected, the main beam steering when rotating the slabs vertical to horizontal states is around 14°. As seen in Fig. 9, the triangular tapered distribution applied is leading to a SLL of about -25 dB in theory. However, by including reflections between the slots in the waveguide, the SLL increased to about -15 dB, which is visible in the measured and simulated SLL in Fig. 17. In larger arrays, slot admittances would be smaller (i.e. cause less reflections) and this should lead to improved SLL.

The cross-polarization levels at H-plane have also been measured. For the operating frequency, they remain 20 dB below the co-polarization level in the main beam direction.

### C. Antenna gain

The realized gain of the antenna for three rotating angles has been measured by the gain comparison method in the anechoic chamber. This gain is shown in Fig. 18a. The antenna efficiency including simulated realized gain and directivity over the frequency band is shown in Fig. 18b. The center frequency shifts in 100 MHz from the design frequency 9.35 GHz to 9.25 GHz for the 45° slabs angle rotation.

The measured reflection coefficient for the 45° slabs angle rotation and at the design frequency is around $-19$ dB, corresponding to a power loss of 1.25%. Based on simulations at the design frequency and for slabs rotation of 45 degrees, the realized gain is 19.52 dB and the directivity is 20.32 dB. This gives an overall antenna efficiency of 83.2%. The sum of return loss ($S_{11} = -19$ dB) and losses in the terminating load ($S_{21} = -11.23$ dB according to Fig. 16), represents 7.53% of power. Therefore 8.02% of the incident power is dissipated in metallic and dielectric materials.

At the design frequency and for the 45° slabs angle rotation, the measured 3-dB gain beamwidth in the H plane is 4.8° which is lower than the simulated value (5.14°). As the gain decreases with scan, the 3-dB gain beamwidth correspondingly increases

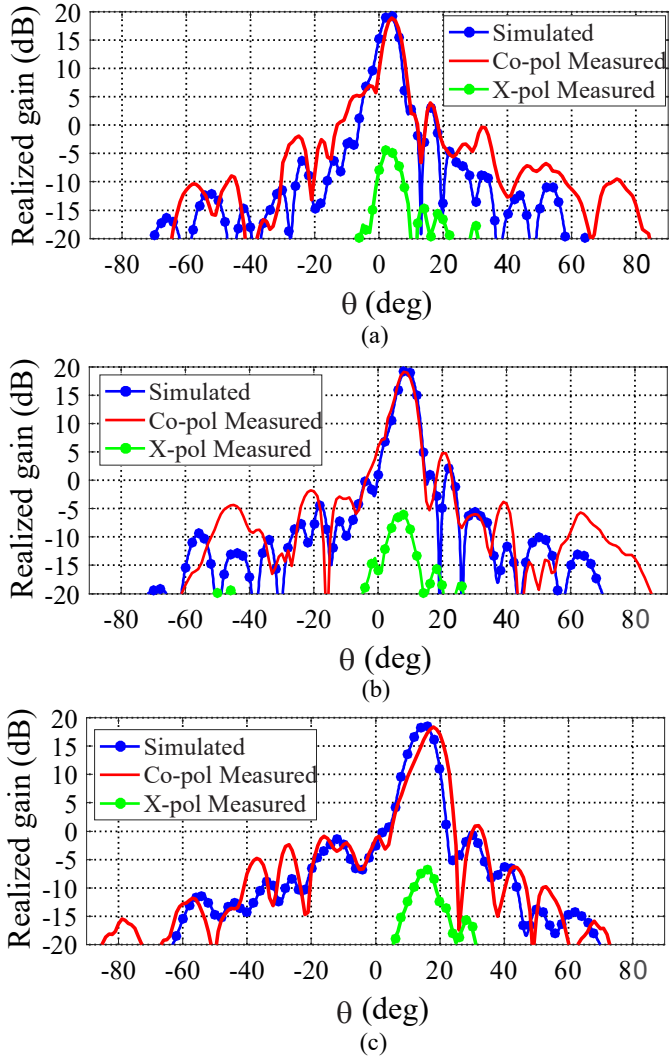

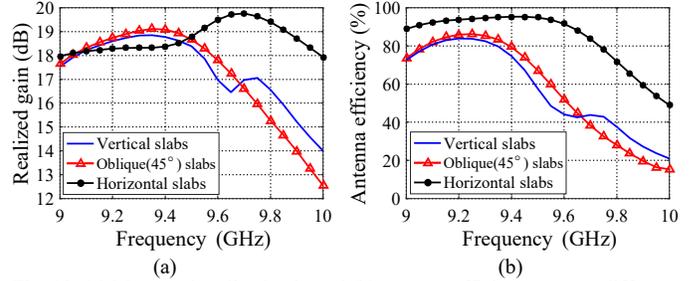

Fig. 18. (a) Measured realized gain and (b) antenna efficiency versus different rotations of the dielectric slabs. The antenna was designed for a frequency of 9.35 GHz and a slabs rotation of 45˚.

TABLE III
H-plane Radiation performance of the antenna at the design at 9.35 GHz

| Slabs rotation states | | Vertical | Oblique (45˚) | Horizontal |
|---|---|---|---|---|
| **Simulated** | Beam direction (˚) | 2.5 | 8.5 | 16.5 |
| | Realized gain (dBi) | 19.59 | 19.52 | 18.51 |
| | 3-dB gain beamwidth (˚) | 5.18 | 5.14 | 7.25 |
| | SLLs (dB) | -16.5 | -16.46 | -18.2 |
| | Realized gain/Directivity (%) | 79.6 | 83.2 | 95 |
| | Dissipated power* (%) | 9.7 | 8.02 | 4.2 |
| **Measured** | Beam direction (˚) | 4 | 8 | 18 |
| | Realized gain (dBi) | 18.85 | 19.11 | 18.33 |
| | 3-dB gain beamwidth (˚) | 5.2 | 4.8 | 6.6 |
| | SLLs (dB) | -14.3 | -15.1 | -17.33 |

*$P_{dissipated} = 1 - |S_{11}|^2 - |S_{21}|^2 - P_{radiated}$

Fig. 17. H-plane radiation patterns $((\phi = 0)$, see the axes presented in Fig.1) of the antenna, simulated and measured in different rotations of the dielectric slabs; (a) Vertical, (b) Oblique (45°) and (c) Horizontal.

to 6.6˚ when the slabs are horizontal. The simulated and measured radiation performance of the antenna in various rotation states of the slabs is summarized in Table III.

## V. CONCLUSIONS

The design of the slot array antenna using rotating dielectric slabs inside the waveguide in order to allow continuous scanning of the antenna main beam has been presented. The antenna shows 14˚ of continuous beam scanning at the design frequency with very good scanning performance. Although the antenna was designed at a single frequency, it presents an overall experimental 10.5% bandwidth for 10 dB return loss. The SLL remains lower than -14 dB over the scanning range. The antenna realized gain ranges between 18.33 dBi to 19.11 dBi. The antenna efficiency including the reflection loss and losses in the terminating load is better than 79% over the scan range.

The choice of a lower permittivity material to realize the dielectric slabs is important. By comparing cases with relative permittivities of 6.15 and 10.2, it was found that in the first case the effect of coupling between the radiating slot elements can be ignored. This has an impact as it simplifies the array design process significantly.

One of the difficulties in the proposed design is to preserve good matching conditions and antenna efficiency for all the rotation states of the slabs. This was achieved by optimizing the tapered transitions between the ridge and dielectric-loaded sections.

Although the experimental evidence of scanning capabilities of rotating a ridge in a metal waveguide slot array has already been explored by Solbach *et al.* [1], the proposed solution has lower side-lobes due to the symmetric field distribution inside the waveguide, more beam scanning range and no TEM mode propagation due to dielectric material used for the slabs.

The presented concept is a very promising candidate for high power continuous beam scanning applications. It could be a less-complicated and cost-effective solution compared to the active phased-array concepts presented in [12-13].

Many antennas presented in literature [16] use periodically loaded leaky waveguides and operate on the frequency scanning principle. In certain applications such as X-band weather radars the allocated bandwidth is very narrow and such concepts cannot be used. The antenna proposed in this paper can realize beam scanning at a fixed frequency, and it is not limited in power handling due to the absence of nonlinear active components. Mechanical scanning speed is of course limited, but remains acceptable for slowly moving targets.

A very good agreement between simulations and measurements also confirms the reliability and accuracy of the slot characterisation previously done. Since no optimization of the complete model has been done, there is still room for improvement of the antenna performance.

## VI. Acknowledgments

The authors would like to thanks the technical staff of Poly-Grames Research Center for the support and the constant interaction during the fabrication and testing phases of this work.

## References

[1] K. Solbach and D. Demirel, "Electro-mechanical beam scanning antenna using rotating ridge inside waveguide slot array," in *2007 2nd International ITG Conf. on Antennas,* Munich, Germany, Mar. 2007.

[2] A. Mirkamali, F. Siaka, J.-J. Laurin, and R. Deban, "Fast and low-cost beam steering using an agile mechanical feed system for exciting circular arrays," *IET Microwaves, Antennas & Propagation,* vol. 10, no. 4, pp. 378–384, Mar. 2016.

[3] V. Manasson, L. Sadovnik, and R. Mino, "MMW scanning antenna," *IEEE AES Syst. Mag.,* pp. 29–33, Oct. 1996.

[4] K. Solbach, R. Schneider, "Review of antenna technology for millimeter-wave automotive sensors", *Proc. Euro. Microw. Conf.*, vol. 11, no. 10, pp. 139–142, Oct. 1999.

[5] R. Schneider and W. J., "High resolution radar for automobile application," *Advances in Radio Science,* vol. 1, no. 6, pp. 105–111, May. 2003.

[6] K. Tekkouk, J. Hirokawa, R. Sauleau, M. Ando, "Wideband and Large Coverage Continuous Beam Steering Antenna in the 60-GHz Band," *IEEE Trans. Antennas Propag.* vol. 65, no. 9, pp. 4418–4426, Sept. 2017.

[7] X. Lu, S. Gu, X. Wang, H. Liu, W. Lu, "Beam-Scanning Continuous Transverse Stub Antenna Fed by a Ridged Waveguide Slot Array," *IEEE Antennas Wireless Propag. Lett.,* vol. 16, pp. 1675–1678, Feb. 2017.

[8] N. K. Host, C.-C. Chen, J. L. Volakis, and F. A. Miranda, "Ku-band traveling wave slot array scanned via positioning a dielectric plunger," *IEEE Trans. Antennas Propag.,* vol. 63, no. 12, pp. 5475–5483, Dec. 2015.

[9] N. K. Host, C.-C. Chen, J. L. Volakis, and F. A. Miranda, "Low cost beam-steering approach for a series-fed array," in *2013 IEEE Int. Symp. on Phased Array Syst. and Technology*, pp. 293–300, Waltham, MA, USA, Oct. 2013.

[10] H. Ward, C. Fowler, and H. Lipson, "GCA radars: Their history and state of development," *Proc. IEEE,* vol. 62, no. 6, pp. 705–716, Jun. 1974.

[11] T. Sabapathy, M. F. B. Jamlos, R. B. Ahmad, M. Jusoh, M. I. Jais, and M. R. Kamarudin, "Electronically reconfigurable beam steering antenna using embedded RF PIN based parasitic arrays (ERPPA)," *Prog. In Electromag. Res.,* vol. 140, pp. 241–261, 2013.

[12] C. Ding, Y. J. Guo, P. Y. Qin, T. S. Bird, Y. T. Yang, "A defected microstrip structure (DMS) based phase shifter and its application in beamforming antennas", *IEEE Trans. Antennas Propag.,* vol. 62, no. 2, pp. 641-651, Feb. 2014.

[13] J. R. De Luis, F. De Flaviis, "A reconfigurable dual frequency switched beam antenna array and phase shifter using PIN diodes", in *2016 IEEE Int. Symp. on Antennas and Propagat. (APSURSI)*, Charleston, SC, USA, June 2009.

[14] E. Ojefors, S. Cheng, K. From, I. Skarin, P. Hallbjorner, and A. Rydberg, "Electrically steerable single-layer microstrip traveling wave antenna with varactor diode based phase shifters," *IEEE Trans. Antennas Propag.,* vol. 55, no. 9, pp. 2451–2460, Sep. 2007.

[15] L. Petit, L. Dussopt, and J.-M. Laheurte, "MEMS-switched parasitic-antenna array for radiation pattern diversity," *IEEE Trans. Antennas Propag.*, vol. 54, no. 9, pp. 2624–2631, Sep 2006.

[16] N. Yang, C. Caloz, and K. Wu, "Full-space scanning periodic phase-reversal leaky-wave antenna," *IEEE Trans. Microw. Theory Tech.,* vol. 58, pp. 2619-2632, Oct. 2010.

[17] A. Ghasemi, J. J. Laurin, "X-band waveguide phase shifter using rotating dielectric slab," in *2016 IEEE Int. Symp. on Antennas and Propagat. (APSURSI)*, Fajardo, Puerto-Rico, June 2016.

[18] M. Navarro-Tapia, J. Esteban, C. Camacho-Penalosa, "On the Actual Possibilities of Applying the Composite Right/Left-Handed Waveguide Technology to Slot Array Antennas," *IEEE Trans. Antennas Propag.* vol. 60, no. 5, pp. 2183–2193, March 2012.

[19] R. E. Collin, *Antennas and Radiowave Propagation*, New York, NY, USA: McGraw-Hill, 1985.